\documentstyle[preprint,aps]{revtex}
\begin{document}
\input{epsf}
\draft
\newfont{\form}{cmss10}
\newcommand{\e}{\varepsilon}
\renewcommand{\b}{\beta}
\newcommand{\unity}{1\kern-.65mm \mbox{\form l}}%
\newcommand{\D}{D \raise0.5mm\hbox{\kern-2.0mm /}}
\newcommand{\A}{A \raise0.5mm\hbox{\kern-1.8mm /}}
\def\pmb#1{\leavevmode\setbox0=\hbox{$#1$}\kern-.025em\copy0\kern-\wd0
\kern-.05em\copy0\kern-\wd0\kern-.025em\raise.0433em\box0}

\def\D{\hbox{\hbox{${D}$}}\kern-1.9mm{\hbox{${/}$}}}
\def\kbar{\hbox{$k$}\kern-0.2truecm\hbox{$/$}}
\def\nbar{\hbox{$n$}\kern-0.23truecm\hbox{$/$}}
\def\pbar{\hbox{$p$}\kern-0.18truecm\hbox{$/$}}
\def\nhbar{\hbox{$\hat n$}\kern-0.23truecm\hbox{$/$}}
\newcommand{\dif}{\hspace{-1mm}{\rm d}}
\newcommand{\dil}[1]{{\rm Li}_2\left(#1\right)}
\newcommand{\diff}{{\rm d}}

\title{Time exponentiation of a Wilson loop for Yang-Mills theories
in $2+\epsilon$ dimensions}
\author{A. Bassetto}
\address{Dipartimento di Fisica ``G.Galilei", Via Marzolo 8, 35131
Padova, Italy\\
INFN, Sezione di Padova, Italy}
\author{R. Begliuomini and G. Nardelli}
\address{Dipartimento di Fisica, Universit\`a di Trento,
38050 Povo (Trento), Italy \\ INFN, Gruppo Collegato di Trento, Italy}

\maketitle
\begin{abstract}
A rectangular Wilson loop centered at the origin, with sides parallel
to space and time directions and length $2L$ and $2T$ respectively,
is perturbatively evaluated ${\cal O}(g^4)$ in Feynman gauge for
Yang--Mills theory in $1+(D-1)$ dimensions. When $D>2$, there is
a dependence on the dimensionless ratio $L/T$, besides the area.
In the limit $T \to \infty$, keeping $D>2$, the leading expression of
the loop involves only the Casimir constant $C_F$ of the fundamental
representation and is thereby in agreement with the expected Abelian-like
time exponentiation (ALTE). At $D= 2$ the result depends also on $C_A$,
the Casimir constant of the adjoint representation and a pure area law
behavior is recovered, but no agreement with ALTE
in the limit $T\to\infty$. Consequences of these results
concerning two and higher-dimensional gauge theories are pointed out.
\end{abstract}
\noindent Padova preprint DFPD 98/TH/32; {\it PACS}: 11.10 Kk, 12.38 Bx

\noindent {\it keywords}: Perturbative Wilson loop calculation; QCD in lower
dimensions
\vfill\eject

\narrowtext

\section{Introduction}
\noindent

There are several reasons why $SU(N)$ Yang-Mills (YM) theories are
interesting in 1+1
dimensions. First of all, this reduction entails tremendous simplifications,
so that many problems can be faced, and often solved, even beyond
perturbation theory. We are referring to exact evaluations of vacuum to vacuum
amplitudes of Wilson loop operators, that, for a suitable choice of contour
and in a particular limit, provide the potential between a static ${\rm q}
{\rm \bar q}$ pair \cite{poly72},\cite{fish},\cite{wils74}.

Another celebrated example is the spectrum of ${\rm q}
{\rm \bar q}$ bound states in the large-N limit,
when dynamical fermions are also taken into account
\cite{hoof74}.

In 1+1 dimensions YM theories without fermions
are considered free theories, apart from topological
effects. This feature looks apparent when choosing an axial gauge.
However they exhibit severe infrared (IR) divergences, which
need to be regularized.
In ref.\cite{hoof74} an explicit IR cutoff is advocated,
which turns out to be uninfluent on the bound state spectrum; a
Cauchy principal value (CPV) prescription in handling the IR singularity
in the gluon propagator leads indeed to the same result \cite{call76}.
On the other hand
this prescription emerges quite naturally if the theory is
quantized at
``equal $x^+$'', namely adopting the light-front $x^+=0$ as quantization
surface.

Still difficulties in performing a
Wick's rotation in the dynamical equations was pointed out
in ref.\cite{Wu}. In order to remedy such a situation, a causal
prescription for the double pole in the
kernel was proposed, which is nothing but the one suggested
years later by Mandelstam and Leibbrandt (ML) \cite{ML},
when restricted to $1+1$ dimensions. This prescription
follows from equal-time quantization\cite{Bas5}
and is mandatory in order to renormalize the theory in 1+3
dimensions\cite{Bas3},\cite{Bas4}.

In view of the above-mentioned results and of the
fact that ``pure'' YM theory does not immediately look free in Feynman gauge,
a test of gauge invariance was performed
in ref.\cite{Bas1} by calculating at ${\cal
O}(g^4)$, both in Feynman and in light-cone gauge with ML prescription,
a rectangular Wilson loop with
light-like sides, directed along the vectors $n_\mu = (T, - T)$,
$n^*_\mu = (L, L)$ and parametrized according to the equations:

\begin{eqnarray}
\label{uno}
C_1 &:& x^\mu (t) = n^{* \mu} t,\nonumber\\
C_2 &:& x^\mu (t) = n^{* \mu} + n^\mu t,\nonumber\\
C_3 &:& x^\mu (t) = n^\mu + n^{* \mu}( 1-t), \nonumber\\
C_4 &:& x^\mu (t) = n^\mu (1 - t), \qquad 0 \leq t \leq 1.
\end{eqnarray}

This contour has been considered in refs.\cite{Korc,Bas2} for
an analogous test of gauge invariance in 1+3 dimensions. Its
light-like character forces a Minkowski treatment.

In order to perform the test, dimensional regularization was used;
the Feynman gauge option is indeed not viable at strictly 1+1 dimensions,
as the usual free vector propagator is not a tempered distribution.

The following unexpected results were obtained.

The ${\cal O}(g^4)$ perturbative loop expression in $d= 1+(D - 1)$
dimensions is finite in the limit $D\to 2$. The results in the two gauges
coincide, as required by gauge invariance. They are function
only of the area $n\cdot n^{*}$ for any dimension $D$ and exhibit
also a dependence
on $C_A$, the Casimir constant of the adjoint representation.

This dependence, when looked at in the light-cone gauge calculation,
comes from non-planar diagrams with the colour factor $C_F(C_F - C_A/2)$,
$C_F$ being the Casimir constant of the fundamental representation.
Besides, there is a genuine contribution proportional to $C_F C_A$
coming from the one-loop correction to the vector propagator. This is
surprising at first sight, as in strictly 1+1 dimensions the
triple vector vertex vanishes in axial gauges. What happens is that
transverse degrees of freedom, although coupled with a
vanishing strength at $D=2$, produce finite contributions when
matching with the self-energy loop singularity precisely at $D=2$,
eventually producing a finite result. Such a
phenomenon could not appear in a strictly
1+1 dimensional calculation, which would only lead to the (smooth)
non-planar diagram result. We stress that this
contribution is essential to get agreement with the Feynman gauge
calculation, in other words with gauge invariance.

We notice that no ambiguity affects our light-cone gauge results,
which do not involve infinities; in addition the discrepancy
cannot be accounted for by a simple redefinition of the coupling,
that would also, while unjustified on general grounds, turn out
to be dependent on the area of the loop.

\smallskip

In order to make the argument complete, we recall that a calculation
of the same Wilson loop in strictly 1+1 dimension in light-cone
gauge with a CPV prescription for the ``spurious'' singularity produces a
vanishing contribution from non-planar graphs. Only planar diagrams
survive, leading to Abelian-like results depending only
on $C_F$, which can be resummed to all orders in the perturbative
expansion to recover the expected exponentiation of the area.

This result, which is the usual one found in the literature,
although quite transparent, {\it does not coincide}
with the Feynman gauge result in the limit $D\to 2$\cite{Bas1}.

The test cannot be generalized to $D\ne 2$ dimensions as
CPV prescription is at odd with causality in this case\cite{Bas4}.

\smallskip

In order to clarify whether there is indeed a pathology in the light-cone
gauge formulation with ML prescription in strictly 1+1 dimensions, one can
try to study the potential $V(2L)$ between a
``static" ${\rm q} {\rm\bar q}$ pair in the fundamental representation,
separated by a distance $2L$. Then a different
Wilson loop is to be considered, {\it viz} a rectangular loop with
one side along the space direction and one side along the time direction,
of length $2L$ and $2T$ respectively. Eventually the limit $T \to \infty$
at fixed $L$ is to be taken: the potential $V(2L)$  between the
quark and the antiquark is indeed
related to the value of the corresponding Wilson loop
amplitude ${\cal W}(L,T)$ through the Abelian-like time exponentiation
(ALTE)
formula \cite{ALTE}
\begin{equation}
\lim_{T\to\infty}{\cal W}(L,T)=const.\  e^{-2i T V(2L)}\ .
\label{potential}
\end{equation}
We remark that this condition is stronger than requiring \cite{fish}
$$\lim_{T\to \infty}{1\over T}\log{\cal W}(L,T)=const.\ ,$$
which, for instance, allows $T$-dependent polynomials factors.

The crucial point to notice in eq.(\ref{potential}) is that 
dependence on the Casimir
constant $C_A$ should cancel at the leading order when $T\to \infty$ in any
coefficient of a perturbative expansion of the potential with
respect to coupling constant. This criterion has often been used
as a check of gauge invariance\cite{Bas4}.

\smallskip

The study of this Wilson loop in strictly 1+1 dimensions has been performed in
light-cone gauge\cite{Bas7}.
In the CPV
case, due to its essentially Abelian nature, the loop can be exactly
evaluated and a simple exponentiation of the area is recovered,
thus providing a
linear potential between the quark and the antiquark. In particular
only the Casimir constant $C_F$ appears.

The corresponding calculation in the ML case
develops genuine non-Abelian terms proportional to $C_A$;
thus, contrary to the previous case, the loop
interaction feels the non-Abelian nature of the theory.
The loop still depends only on the area, but
ALTE is lost:
in the limit $T \to \infty$ a dependence on
$C_A$ survives \cite{Bas7},\cite{Stau}.

Although ALTE can be proven
only introducing the ``true spectrum'' of the theory, which
is beyond the perturbative regime, we feel
somehow uneasy with this conclusion, which is at variance
with all the previous tests of gauge invariance considered in the
literature. We are thereby led to consider this phenomenon as a peculiarity
of strictly two-dimensional theories, which occurs when topological excitations
are disregarded as in perturbative loop calculations.

Then the interest arises in performing the same Wilson loop
calculation in dimensions $D>2$. It may indeed happen that
the limits $T\to \infty$ and $D\to 2$ do not commute.

The main result of this paper, namely that,
in the large $T$-limit when $D>2$, the perturbative ${\cal O}(g^4)$
contribution is in agreement with ALTE, supports
the conclusion above.

\smallskip

Owing to the very heavy
character of the calculation, we choose to work in Feynman gauge.
In Sect.2 we review basic concepts and notation. We also compute
the Wilson loop at ${\cal O}(g^2)$. In Sect.3
the relevant diagrams at ${\cal O}(g^4)$ are computed in $D=2\omega$
dimensions, after collecting them in three basic families, the
``exchange'' diagrams, the ``self-energy'' and the ``spider''
ones, namely those involving the triple vector vertex.
For $D>2$ the loop depends, besides on the area,
also on the ratio $\beta=L/T$. As we are eventually interested 
in the large-$T$ behavior, we shall always consider the region $\beta<1$;
moreover we shall choose $\omega=1+\epsilon$, with a small, positive
$\epsilon .$
Sect.4 contains our conclusions. In it we show that,
when $D>2$, agreement with ALTE occurs
in the limit $T\to \infty$, with a pure $C_F$-dependence in the
leading coefficient. Agreement with all previous
results \cite{Bas4} in higher dimensions is thus re-established.

To perform instead the limit $D\to 2$ for generic
values of $\beta$ in a rigorous way, turns out
to be extremely difficult from a technical viewpoint. As a matter of
fact the contribution coming from graphs with the triple gluon
vertex, exhibits quite involved analyticity properties in the
variable $\beta$. Nevertheless, if such a limit is performed accepting
a natural conjecture we shall explain later on, the result one
gets exactly coincides with the gauge invariant one obtained in
ref.\cite{Bas1} for a loop of the same area with light-like sides;
thereby we enforce the argument that in two dimensions a pure
area behavior is to be expected, no matter the orientation and the
shape of the loop. In so doing, however, in two dimensions at 
${\cal O}(g^4)$ a $C_A$-dependence is
definetely there and there is no agreement with ALTE.

The theory looks indeed discontinuous in the limit $D\to 2$.

\smallskip

Details of the calculations are given in the Appendices.

\section{General considerations}
\noindent
We consider the usual ``gauge fixed'' Yang-Mills
Lagrangian,
\begin{equation}
{\cal L} = -{1\over 4} F^a_{\mu \nu} F^{a\mu \nu}
+{1\over 2} \partial^{\mu}A_{\mu}^{a} \partial^{\nu}A_{\nu}^{a} .
\label{lagrangian}
\end{equation}
Without loss of generality, we consider $SU(N)$ as gauge group, so that
the field strength in (\ref{lagrangian}) is  defined as $F_{\mu \nu}^a=
\partial_\mu A_\nu^a - \partial_\nu A_\mu^a + g f^{abc} A_\mu^b A_\nu^c$,
$f^{abc}$ being the structure constants of $SU(N)$.

\smallskip

We consider the closed path $\gamma$
parametrized by the following  four segments
$\gamma_i$,
\begin{eqnarray}
\gamma_1 &:& \gamma_1^\mu (s) = (sT, L)\ ,\nonumber\\
\gamma_2 &:& \gamma_2^\mu (s) = (T,-sL)\ ,\nonumber\\
\gamma_3 &:& \gamma_3^\mu (s) = (-sT, -L)\ , \nonumber\\
\gamma_4 &:& \gamma_4^\mu (s) = (-T, sL)\ , \ \ \qquad -1 \leq s \leq 1.
\label{path}
\end{eqnarray}
describing a  (counterclockwise-oriented) rectangle
centered at the origin of the plane ($x^1,x^0$),
with length sides $(2L,2T)$, respectively (see Fig. \ref{fig1}).
Then, for the definition of the Wilson loop around $\gamma$ we shall adopt the
standard one, given by the following  vacuum to vacuum amplitude
\begin{equation}
{\cal W}_\gamma (L,T) = {1\over N} \langle 0| {\rm Tr}\left[ {\cal T}{\cal P}
{\rm exp} \left( ig \oint_\gamma dx^\mu \ A^a_\mu (x) T^a \right)\right]
|0\rangle \ \ , \label{wilson}
\end{equation}
where ${\cal T}$ orders gauge fields in time and ${\cal P}$ orders generators
$T^a$ of the gauge group $SU(N)$ along the closed integration path $\gamma$.
The perturbative expansion of the Wilson loop (\ref{wilson}) looks like
\begin{equation}
{\cal W}_\gamma (L,T) = 1 + {1\over N}\sum_{n=2}^\infty (ig)^n \oint_\gamma
dx_1^{\mu_1} \cdots \oint_\gamma dx_n^{\mu_n}\theta( x_1 >\cdots >x_n )
{\rm Tr} [ G_{\mu_1 \cdots \mu_n} (x_1,\cdots ,x_n)]\ ,
\label{wilpert}
\end{equation}
where $ G_{\mu_1 \cdots \mu_n} (x_1,\cdots ,x_n)$ is the Lie algebra valued
$n$-point Green function, in which further dependence on the coupling
constant is usually buried; the Heavyside $\theta$-functions order the points
$x_1,\cdots ,x_n$ along the integration path $\gamma$.

It is easy to show that the perturbative expansion of ${\cal W}_\gamma$ is an
even power series in the coupling constant, so that we can write
\begin{equation}
{\cal W}_\gamma (L,T)= 1+g^2 {\cal W}_2 + g^4 {\cal W}_4 + {\cal O}(g^6)\ .
\label{pert}
\end{equation}

An explicit evaluation of the function  ${\cal W}_2$ in eq. (\ref{pert})
gives the diagrams contributing to the loop with a single exchange (i.e. one
propagator), namely
\begin{equation}
{\cal W}_2= - {1\over 2} C_F \oint \oint D_{\mu\nu} (x-y) dx^\mu dy^\nu \ ,
\label{w2}
\end{equation}

where $D_{\mu\nu} (x)$ is the usual free propagator in Feynman gauge
$$D_{\mu\nu} (x)=-g_{\mu\nu}{{{\pi}^{-D/2}}\over 4}\Gamma(D/2-1)
(-x^2+i\epsilon)^{1-D/2}.$$

A fairly easy calculation leads to the result
\begin{eqnarray}
\label{singleexchange}
{\cal W}_2&=&{{C_F}\over {{\pi}^{\omega}}}(2L)^{2-2\omega} LT \Big[i
\Gamma(\omega-3/2) \Gamma(1/2)\\
&+&{{2\beta\Gamma(\omega)}\over {\omega -2}}\Big({1\over {3-2\omega}}-
e^{-i\pi\omega}\sum_{n=1}^{\infty}{{\Gamma(n+\omega-2)}\over {\Gamma(\omega
-2)}}{{\beta^{2n+2\omega-4}}\over {(2n-1)(2n+2\omega-3) n!}}\Big)\Big].
\nonumber
\end{eqnarray}
It should be noticed that this result does not coincide with the corresponding
one in ref. \cite{Bas1}, which was evaluated in the same gauge choice but with
the loop sides along the  $x^+$ and $x^-$ directions. Contrary to what happens
in ref. \cite{Bas1}, eq. (\ref{singleexchange}) exhibits an explicit
dependence on the
ratio $\beta=L/T$. Only in the two dimensional limit the two results coincide:
the limit $\omega \to1$ is smooth and gives the pure area dependence $${\cal
W}_2=-2iC_FLT.$$

\bigskip

The diagrams contributing to
${\cal W}_4$ can be grouped into three distinct families

\noindent ${\cal W}_4={\cal W}^{(1)}+{\cal W}^{(2)}+{\cal W}^{(3)}$:

\begin{itemize}
\item the ones with a double gluon exchange in which the propagators
can either cross or uncross;
\item the ones in which the gluon propagator contains a self-energy
correction;
\item the ones involving a triple vector coupling.
\end{itemize}

\noindent
In strictly two dimensions and in an axial gauge
only the first family is present. In Feynman gauge all of them must
be taken into account; moreover we are here considering the problem
in D-dimensions. The first family is also the only one contributing
to the Abelian case.

\section{Wilson loop calculations}
\noindent

We start by considering the diagrams belonging to the first category:
a straightforward calculation gives
\begin{equation}
{\cal W}^{(1)}={1\over 8N}\oint\oint\oint\oint{\rm Tr}[{\cal P}
 (T^a_xT^a_yT^b_z T^b_w)]
  D_{\mu\nu}(x-y)D_{\rho\sigma}(z-w)\;
 dx^{\mu}dy^{\nu}dz^{\rho}dw^{\sigma}\;\; ,
\label{w4}
\end{equation}
where subscripts in the matrices have been introduced
to specify their ordering.
From eq. (\ref{w4}), the diagrams with two-gluons exchanges contributing to the
order $g^4$ in the perturbative expansion of the Wilson loop fall into two
distinct classes, depending on the topology of the diagrams:
\begin{enumerate}
\item {\it Non-crossed diagrams}: if the pairs $(x,y)$ and $(z,w)$
are contiguous around the loop the two propagators do not cross (see Fig.
\ref{fig2}a) and the trace in (\ref{w4}) gives ${\rm Tr} [T^aT^aT^bT^b] = N
C_F^2$.
\item {\it Crossed diagrams}: if the pairs $(x,y)$ and $(z,w)$
are  not contiguous around the loop the two propagators do cross (see
Fig. \ref{fig2}b) and the trace in (\ref{w4}) gives ${\rm Tr} [T^aT^bT^aT^b] =
{\rm Tr}[ T^a (T^a T^b + [T^b,T^a])T^b]=
N (C_F^2 - (1/2) C_A C_F)$, $C_A$ being
the Casimir constant of the adjoint representation defined by
$f^{abc}f^{d bc}= C_A \delta^{ad}$.
\end{enumerate}
We see that  the $C_F^2$ term is present in both types of diagrams and with
the same coefficient. This term is usually denoted  ``Abelian term'': were the
theory Abelian, only such $C_F^2$ terms would contribute to the loop. On the
other hand, the $C_FC_A$ term is present only in crossed diagrams, and is
typical of non-Abelian theories.

Thus, we can decompose  ${\cal W}^{(1)}$ as the sum of an Abelian and a
non-Abelian part,
\begin{equation}
{\cal W}^{(1)}={\cal W}_{ab} +{\cal W}_{na}\ \ .
\end{equation}
 Moreover,  the Abelian part is simply half of the square of the
order-$g^2$ term, i.e.
\begin{eqnarray}
{\cal W}_{ab} &= & {1\over 8} C_F^2
\oint\oint\oint\oint
       D_{\mu\nu}(x-y)D_{\rho\sigma}(z-w)\;
        dx^{\mu}dy^{\nu}dz^{\rho}dw^{\sigma}\nonumber\\
&=& {1\over 2} \left( - {1\over 2} C_F \oint \oint D_{\mu\nu} (x-y)
dx^\mu dy^\nu \right)^2 \ \ .\label{w4ab}
\end{eqnarray}
Equation (\ref{w4ab}) is just a particular case
of a more general theorem due to
Frenkel and Taylor \cite{Fre}, which proves that
the only relevant
terms in the perturbative expansion of the loop are the so-called ``maximally
non-Abelian'' ones; at ${\cal O}(g^4)$ those terms are just proportional to
$C_FC_A$. They are the only ones we have to evaluate to get ${\cal W}_{na}$.

All the Abelian  terms
(depending only on $C_F$) in
the perturbative expansion of the Wilson loop sum up to reproduce the Abelian
exponential
\begin{equation}
{\cal W}_\gamma^{ab} (L,T) = {\rm exp}\left( - {1\over 2} C_F g^2
\oint\oint D_{\mu \nu}(x-y) dx^\mu dy^\nu \right) \  , \label{abexp}
\end{equation}
where the result in eq.(\ref{singleexchange}) can be introduced.
This result at ${\cal O}(g^4)$ has been explicitly checked calculating
the relevant non-crossed exchange diagrams (see Appendix A for details).

In the limit $D\to 2$ the simple exponentiation of the area is easily
recovered
\begin{equation}
{\cal W}_\gamma^{ab} (L,T)=exp [-{i \over 2}C_F g^2{\cal A}],
\label{area}
\end{equation}
where ${\cal A}=4LT$ is the area of the loop.

\bigskip

We have now to calculate  loop
integrals of the type given in eq.(\ref{w4}). In view of
the parametrization  (\ref{path}), it is convenient to decompose loop integrals
as sums of integrals  over the segments $\gamma_i$, and to this purpose we
define
\begin{equation}
E_{ij}(s,t) = D_{\mu \nu}\bigl[ \gamma_i(s) -\gamma_j(t) \bigr]
\dot\gamma_i^\mu(s) \dot\gamma_j^\nu(t) \ , \qquad i,j=1,\dots ,4\ \, ,
\label{e}
\end{equation}
where the dot denotes the
derivative with respect to the variable parametrizing the
segment. In this way, each diagram can be written as integrals of products of
functions of the type (\ref{e}).  Each graph will be labelled by
a set of pairs
$(i,j)$, each pair denoting a gluon propagator joining the segments $\gamma_i$
and $\gamma_j$.

Due to the symmetric choice of the contour $\gamma$ and to the  fact that
propagators are even functions, i.e. $D_{\mu
\nu}(x)=D_{\mu \nu}(-x)$, we have the following identities that halve the
number of diagrams to be evaluated:
\begin{eqnarray}
E_{ij}(s,t) &=& E_{ji}(t,s)\ ,\nonumber\\
E_{11}(s,t) &=& E_{33}(s,t)\ ,\nonumber\\
E_{22}(s,t) &=& E_{44}(s,t)\ .
\label{sym}
\end{eqnarray}

We remind the reader that in Feynman gauge the propagators can attach
either to the same rectangle side or to opposite sides, but not on a
couple of contiguous ones.

We have now to consider the ${\cal O}(g^4)\  C_F C_A$-terms. From the previous
discussion it follows that only ``crossed
diagrams'' (maximally non-Abelian ones)  need to be evaluated

\begin{eqnarray}
{\cal W}_{na}& =& - {1\over 2} C_AC_F
\sum_{i,j,k,l}{}^{'} \int ds\int dt\int du\int dv \
E_{ij}(s,t)E_{kl}(u,v)\nonumber\\ &\equiv&- {1\over 2} C_AC_F
\sum_{i,j,k,l}{}^{'} C_{(ij)(kl)}\ ,
\label{doppi}
\end{eqnarray}
where the primes mean that we have to sum only over
crossed propagators configurations and over topologically inequivalent
contributions, as carefully explained in the following;
we have not specified the integration
extrema as they depend on the particular type of crossed diagram we are
considering (the extrema
must be chosen in such a way that propagators remain crossed).

The last equality in eq. (\ref{doppi}) defines
the general diagram $C_{(ij)(kl)}$: it is a  diagram with two {\it crossed}
propagators joining the sides $(ij)$ and $(kl)$ of the contour (\ref{path}).
In Fig. \ref{fig3} a few examples of diagrams are drawn to get the reader
acquainted with the
notation.
 The first of eq.
(\ref{sym})  permits to select just 11 types of topologically distinct
crossed diagrams.
The remaining symmetry
relations (\ref{sym}) further lower the number to 7.
As a matter of fact, although topologically inequivalent, from eq. (\ref{sym})
it is easy to get
\begin{eqnarray}
C_{(11)(11)}&= C_{(33)(33)}\ \ , \qquad C_{(22)(22)}=& C_{(44)(44)}\
\ ,\nonumber\\
C_{(11)(13)}&= C_{(33)(13)}\ \ , \qquad C_{(22)(24)}=& C_{(44)(24)}.
\label{relations}
\end{eqnarray}
which are the 4
relations needed to lower the number of diagrams to be evaluated
from 11 to 7. Besides the 8 diagrams quoted in eq. (\ref{relations}), there
are three other
crossed diagrams that do not possess any apparent symmetry relation
with other diagrams: $C_{(13)(13)}, \ C_{(24)(24)}$ and $C_{(13)(24)}$ (see
Fig. \ref{fig4}), so that the number of topologically inequivalent crossed
diagrams is indeed 11.

The
calculation of the 7 independent  diagrams needed is lengthy and  not trivial.
The details of such calculation are
reported in  Appendix A. Each diagram depends not only on the area
$A=4LT$ of the loop, but also on the dimensionless ratio $\beta = L/T$ through
complicated multiple integrals involving powers depending on $\omega$.

Adding all the contributions as in eq. (\ref{doppi}) we eventually
arrive at the
following result ${\cal O}(g^4)$ for the non-Abelian part of the exchange
diagrams contribution:
\begin{eqnarray}
\label{wmlg4}
 {\cal W}_{na}
&=&C_F C_A{{(2T)^{4-4\omega}}\over {\pi^{2\omega}}}(LT)^2 e^{-2i\pi\omega}\\
&&\times\Big({{\Gamma^2(\omega-1)}\over {(2\omega-4)(2\omega-3)}}\Big[
1+{{1-\omega}\over {(4\omega-5)(2\omega-3)}}+{\cal O}(\beta^{5-4\omega})
\Big]\Big).
\nonumber
\end{eqnarray}

We notice that the expression above exhibits a double and a single
pole at $\omega=1$, whose Laurent expansion gives
\begin{equation}
\label{crocilaurent}
{{\cal W}_{na}\pi^{2\omega} e^{2i\pi  \omega}\over C_FC_A (2T)^{4-4\omega}
 (LT)^2}={1\over 2(\omega-1)^2}+{1-\gamma\over(\omega -1)} -1-2\gamma+\gamma^2
+{\pi^2\over 12} +{\cal O}(\omega -1)\ ,
\end{equation}
$\gamma$ being the Euler-Mascheroni constant.
\vskip .5truecm

We turn now our attention to the calculation of ${\cal W}^{(2)}$,
namely of the diagrams with a single gluon exchange in which the
propagator contains a self-energy correction ${\cal O}(g^2)$.
Of course both gluon and ghost contribute to the self-energy.
The color factor is obviously a pure $C_F C_A$.

We call them ``bubble'' diagrams. We denote by $B_{ij}$ the
contribution of the diagram in which the propagator connects the
rectangle segments $\gamma_i$, $\gamma_j$ (see Fig. \ref{fig5}).

\smallskip

There are 10 topologically inequivalent diagrams; however, the symmetries
we have already discussed and the symmetric choice of the contour
entails the four conditions
$B_{11}=B_{33}$, $B_{22}=B_{44}$, $B_{12}=B_{34}$, $B_{14}=B_{23}$,
whereas the remaining two diagrams $B_{13}$ and $B_{24}$ are unrelated by any
symmetry relation. In addition, it is easy to see by perfoming a simple change
of variable that $B_{14}$ and $B_{34}$ are equal. Thus, there are
5 independent diagrams to be evaluated
\begin{equation} \label{bubblelist} {\cal
W}^{(2)}=B_{13}+B_{24}+2B_{11}+2B_{22}+4B_{12}.
\end{equation}

The details of the calculation are described in Appendix   B.
We here report the final result in the form of an expansion with
respect to the variable $\beta$
\begin{eqnarray}
\label{bubble}
{\cal W}^{(2)}&=&C_FC_A{{(2T)^{4-4\omega}}\over {\pi^{2\omega}}}(LT)^2
e^{-2i\pi\omega} \times\\ \nonumber
&&\Big[{{(3\omega -1) \Gamma^2(\omega)}\over {2\Gamma(2\omega)
\Gamma(4-\omega)}}\Gamma(1-\omega)\Gamma(2\omega -2)\Big(
{{2\omega -6}\over {5-4\omega}}+{\cal O}(\beta^{5-4\omega})\Big)\Big].
\end{eqnarray}
Again we notice the presence of a double and of a single pole at $\omega =1.$
The relevant Laurent expansion is
\begin{equation}
{{\cal W}^{(2)}\pi^{2\omega} e^{2i\pi  \omega}\over C_FC_A (2T)^{4-4\omega}
 (LT)^2}={1\over (\omega-1)^2}+{9-4\gamma\over2(\omega -1)}+{39\over 2}
-9\gamma +2\gamma^2 +{\pi^2\over 6}
  +{\cal O}(\omega -1)\ ,
\label{bollelaurent}
\end{equation}
\vskip .5truecm

The third quantity ${\cal W}^{(3)}$ is by far the most difficult one
to be evaluated. It comes from ``spider'' diagrams, namely the diagrams
containing the triple gluon vertex (see Fig. \ref{fig6}).
We denote by $S_{ijk}$ the contribution of the diagram in which the
propagators are attached to the segments $\gamma_i$, $\gamma_j$, $\gamma_k$.

It can be checked that all the spiders with the three legs attached to the same
line vanish, as well as the spiders with two legs on one side and the third leg
attached to the opposite side, {\it i.e.} $S_{111}= S_{222}=S_{333}=S_{444}=
S_{113}=S_{133}=S_{224}=S_{244}=0$. Thus, there
are  12 non-vanishing topologically inequivalent diagrams; however, their number
is just halved by the symmetric choice of the contour ($S_{124}=S_{234}$,
$S_{123}=S_{134}$, $S_{112}=S_{334}$,
$S_{233}=S_{114}$, $S_{344}=S_{122}$, $S_{144}=S_{223}$), so that ${\cal
W}^{(3)}$ can be expressed in terms of the remaining 6 independent ones
\begin{equation} \label{spiderlist}
{\cal W}^{(3)}=2S_{112}+2S_{123}+2S_{124}+2S_{233}+2S_{144}+2S_{344}.
\end{equation}
Each term is represented by a multiple integral which cannot be
evaluated for a generic dimension $\omega$ in closed form. In
particular it exhibits complicated analyticity properties in the
variable $\beta$, just in the neighborhood of the value $\beta=0$
which is of interest for us. Details of the calculation are again deferred
to an Appendix (Appendix  C).

We have succeeded in obtaining the following result for $D>2$
\begin{equation}
\lim_{\beta \to 0}{{\cal W}^{(3)}\pi^{2\omega} e^{2i\pi  \omega}
\over C_FC_A (2T)^{4-4\omega}
 (LT)^2}=-{3\over 2(\omega-1)^2}+{3\gamma -11/2\over(\omega -1)}
-{35\over 2} +11\gamma -3\gamma^2 +{\pi^2\over 12}
 +{\cal O}(\omega -1)\ .
\label{ragnilaurent}
\end{equation}
A double and a single pole at $D=2$ again are present in this expression.

\section{Concluding remarks}
\noindent
Since the Abelian part of our results depends only on the
Casimir constant $C_F$ and smoothly exponentiates in the large-$T$
limit even when $D>2$ (see eqs.(\ref{singleexchange},\ref{abexp})), in
the following
we focus our attention on the non-Abelian part, namely on the quantity
containing the factor $C_F C_A$  $${\cal W}_{na}+
{\cal W}^{(2)}+{\cal W}^{(3)}.$$

It is actually convenient to divide by the square of the loop area,
by introducing the new expression ${\cal N}$
\begin{equation}
\label{rid}
{\cal N} C_F C_A (LT)^2={\cal W}_{na}+
{\cal W}^{(2)}+{\cal W}^{(3)}.
\end{equation}
Then, from eqs.(\ref{wmlg4},\ref{bubble},\ref{ragnilaurent}), it is easy to
conclude that, thanks to the factor $T^{4-4\omega}$, ${\cal N}$
vanishes in the limit $T\to \infty$ when $\omega>1$.

This is precisely the usual necessary condition required
at ${\cal O}(g^4)$ in order to
get agreement with ALTE when summing higher orders\cite{Bas4}.

We cannot discuss the limit $\omega \to 1$ for generic (small) values
of $\beta$ in our results as we are
only able to master the expressions at $\beta=0$.

Nevertheless if we consider the quantity
$$\lim_{\omega \to 1}\lim_{\beta \to 0} (T^{4\omega -4}{\cal N})$$
we get a quite interesting result. Indeed double and simple poles
at $\omega =1$ cancel in the sum, leading to
\begin{equation}
\label{dibiagio}
\lim_{\omega \to 1}\lim_{\beta \to 0} (T^{4\omega -4}{\cal N})=
{1\over {\pi^2}}(1+{{\pi^2}\over 3}),
\end{equation}
which {\it exactly} coincides with eq.(13) of \cite{Bas1}.

This is, first of all, a formidable check
of all our calculations, if we conjecture that the same result
is obtained by performing the limit $$\lim_{\beta \to 0}
\lim_{\omega \to 1} {\cal N}.$$ But it also entails the quite non trivial
consequences we are going to discuss.

At a first sight one could have thought to recover the result
obtained in strictly 1+1 dimensions in ref.\cite{Bas7}. This is not the case;
as a matter of fact in strictly 1+1 dimensions and in light-cone gauge
the contribution from diagrams containing a self-energy insertion
is missing, in spite of the fact that, if calculated first at $\omega>1$,
it does not vanish in the limit. This phenomenon has been discussed
at length in ref.\cite{Bas1}.

What may be surprising is that this extra contribution, required by gauge
invariance, when considered in the limit $\omega\to 1$, exhibits a pure
area dependence on its own, being the same no matter the orientation of
the loop. Namely, the geometrical arguments (invariance under
area-preserving diffeomorphism) which lead to a pure area
dependence in two dimensions, but not in higher ones, are recovered in
the limit $\omega\to 1$, in spite of the singular nature of this limit,
and of the difference in the two results (with and without self-energy
diagrams).

We consider this point quite intriguing; it seems that, in order to
get beyond two dimensions towards higher ones, the theory
needs further inputs which cannot be {\it a priori} guessed in two
dimensions. On the other hand it is known that operators exist which
are irrelevant at $\omega>1$, but can be competitive in exactly two dimensions
\cite{doug}.

Our result can therefore be interpreted as a warning when one tries to
extend straightforwardly conclusions obtained in strictly two dimensions
to more realistic situations.

Before concluding let us summarize the main perturbative features 
which are common to both Feynman and light-cone gauges.

For $D>2$ the ${\cal O}(g^4)$ result depends on the contour: if it has 
lightlike sides, there is a dependence only on the area, no matter the
value of $D$. If instead the contour is a space-time rectangle, for $D>2$
there is also a dependence on the ratio $\beta \equiv L/T$, still
reproducing the expected exponential behaviour in the limit $T \to \infty
\ (\beta =0).$

The limit $D \to 2$ of such ${\cal O}(g^4)$ result is finite and depends only
on the area, no matter the orientation and the shape of the contour. It
consists of two addenda (compare eq.(27) with eq.(13) of \cite{Bas1}). 

In light cone gauge the second addendum comes from `crossed' diagrams, 
the first one is due to the
self-energy correction to the gluon propagator. In axial gauges the transverse
degrees of freedom are coupled with a strenght of order $D-2$; nevertheless
they produce a finite contribution when matching the self-energy loop
singularity precisely at $D=2$.

In exactly 1+1 dimensions therefore the 
first addendum is missing and one just gets the result of refs.\cite{Bas7} 
and \cite{Stau}.
Its area dependence is hardly surprising in view of the symmetry under
area preserving diffeomorphisms at $D=2$; it is perhaps remarkable that also
the first addendum, which originates at $D>2$, exhibits in the limit
$D \to 2$ a pure area dependence on its own. 
However
both contributions contain the factor $C_F C_A$ and thereby disagree
with the simple area exponentiation.
At a perturbative level there is a discontinuity, which 
is not surprising in the light of the arguments presented in ref.\cite{Stau};
however it does not explain why agreement with the simple area behaviour 
is not recovered.

Moreover, the authors of ref.\cite{Stau}, working in exactly 1+1 dimensions,
have succeeded in resumming the perturbative series: the result they get 
does not exhibit the usual purely exponential area law.
Besides the appearance of Laguerre polynomials, the coefficient in the
exponent multiplying the area is different from the usual one.
We are thereby faced with a discrepancy.

\smallskip

Now we want to discuss the consequences concerning ALTE
in the large-$T$ limit. We have checked that our findings at ${\cal O}
(g^4)$ comply with ALTE, as long as $\omega>1$.
At $\omega=1$ this is not the case; however it is likely that ALTE
in this case is restored by genuine non perturbative contributions,
reconciling gauge invariance with basic spectral properties and
solving the paradox.

As a matter of fact
the usual $D=2$ well established result can most safely be obtained
by means of non perturbative techniques. Actually one could also get it
by resumming the perturbative series, but only in axial gauges with
a peculiar prescription for the propagator (the Cauchy principal
value, which is doubtful in $D=2$ and certainly incorrect for
$D>2$). The reason why this happens might be deep 
and related to peculiar properties of the $D=2$ light-front vacuum 
in light-front quantization.

At a non-perturbative level there are arguments
which show that at large $N$ and on the sphere $S_2$ a phase
transition occurs, induced by instantons (see ref.\cite{boul}). The purely
exponential area behaviour occurs only in the strong coupling 
phase, which is 
dominated by instantons. In the weak coupling phase the result is
completely different and agrees (after decompactification) with the
ones of refs.\cite{Bas7} and \cite{Stau}. This is the reason why we 
believe that the discrepancy can be solved by taking instantons into
account \cite{Basgri}; they obviously do not contribute to any perturbative
calculation, even when the perturbative series is fully resummed.

\smallskip

One could say that perturbative
results are irrelevant at $D=2$. We cannot share this opinion.
At $D>2$ perturbative results are the only easily accessible ones;
several perturbative tests of gauge invariance have been performed in the past
at $D>2$, just {\it assuming}
the simple area exponentiation in the large $T$-limit.
One of the results 
of this paper is to prove that indeed at $D>2$ such an exponentiation 
occurs as expected.

We do not think it is immaterial to understand how these nice features behave
in the transition $D \to 2$, also in order to contrast them against
genuine non perturbative results.

Why genuinely non perturbative contributions, which are likely to be
relevant also in higher dimensions, are crucial only in two dimensions
to possibly recover ALTE, is at present unknown.
This very interesting issue is under investigation and will be discussed
elsewhere.

\vfill
\eject

\appendix
\section{exchange diagrams}
In order to get the ${\cal O}(g^4)$ contribution to the Wilson Loop arising
from the exchange diagrams, we only have to evaluate the maximally non Abelian
diagrams, that in the present case are the so called crossed diagrams. In fact,
the contribution coming from planar diagrams
can be easily obtained through the Abelian exponentiation theorem, as explained
in the main text. In this appendix we shall give the main sketch of computation
of the seven independent crossed diagrams needed:

\begin{eqnarray}
\label{crociapp1}
C_{(13)(13)}&=&\!\!\int_{-1}^{1} \!\!du\int_{-1}^{1} \!\!dv {\alpha T^4\over[4
L^2 - (u+v)^2 T^2 + i \epsilon]^{\omega -1}} \int_u^1\!\! dt \int_v^1 \!\!ds
{1\over [4 L^2 - (t + s)^2 T^2 + i \epsilon]^{\omega - 1}},\\
C_{(24)(24)}&=&\!\!\int_{-1}^{1} \!\!du\int_{-1}^{1} \!\!dv {\alpha L^4\over[L^2
(u+v)^2- 4T^2 + i \epsilon]^{\omega -1}} \int_u^1\!\! dt \int_v^1 \!\!ds {1\over
[ L^2 (t+s)^2 - 4 T^2 + i \epsilon]^{\omega - 1}},\\
C_{(13)(24)}&=&\!\!\int_{-1}^{1} \!\!du\int_{-1}^{1} \!\!dv {(-
\alpha L^2T^2)\over[ L^2 (u+v)^2 - 4 T^2 + i \epsilon]^{\omega -1}}
\int_{-1}^1\!\! dt \int_{-1}^1 \!\!ds {1\over [4 L^2 - (t + s)^2 T^2 + i
\epsilon]^{\omega - 1}},\\
C_{(11)(13)}&=&\!\!\int_{-1}^{1} \!\!du\int_{-1}^{1}
\!\!dv {(-2\alpha T^4)\over[4 L^2 - (u+v)^2 T^2 + i \epsilon]^{\omega -1}}
\int_{-1}^u\!\! dt \int_u^1 \!\!ds {1\over [ - (t - s)^2 T^2 + i
\epsilon]^{\omega - 1}},\\
C_{(22)(24)}&=&\!\!\int_{-1}^{1} \!\!du\int_{-1}^{1}
\!\!dv {(-2\alpha L^4)\over[ L^2 (u+v)^2- 4 T^2 + i \epsilon]^{\omega -1}}
\int_{-1}^u\!\! dt \int_u^1 \!\!ds {1\over [(t - s)^2 L^2 + i \epsilon]^{\omega
- 1}},\\
 C_{(11)(11)}&=&\!\!\int_{-1}^{1} \!\! dt\int_{t}^{1} \!\! du \int_{u}^1
\!\! ds\int_s^1 \!\! dv {(2\alpha T^4)\over [ - (t-s)^2 T^2 + i\epsilon]^{\omega
- 1} [ - (u - v)^2 T^2 + i \epsilon]^{\omega -1}},\\
C_{(22)(22)}&=&\!\!\int_{-1}^{1} \!\! dt\int_{t}^{1} \!\! du \int_{u}^1 \!\!
ds\int_s^1 \!\! dv {(2\alpha L^4)\over [  (t-s)^2 L^2 + i\epsilon]^{\omega - 1}
[  (u - v)^2 L^2 + i \epsilon]^{\omega -1}},
\end{eqnarray}
where  $\alpha =[\Gamma (\omega-1)]^2/(16\pi^{2\omega})$.
 As an example,
we
report the main sketch of computation of $C_{(13)(13)}$.

In order to perform the integrations a series expansion of the
denominators in (A1) is not enough as the series do not converge in the
entire integration domains. The necessary analytic continuations
are provided by
two Mellin-Barnes transformations:

\begin{eqnarray}
\label{MB1}
C_{(13)(13)}&=& \alpha T^4 (4 L^{2})^{2 - 2 \omega}
{1 \over{[\Gamma(\omega-1)]^2}} {1 \over {(2 \pi i)^2}}
\int_{-i \infty}^{+i
\infty} dy \Gamma(\omega -1 + y) \Gamma(-y) \left(
{1 \over{- \beta^2 - i
\epsilon}}\right)^y  \times \nonumber \\ &&\int_{-i \infty}^{+i \infty} dz
\Gamma(\omega
-1 + z) \Gamma(-z) \left( {1 \over{- \beta^2 - i \epsilon}}\right)^z \times
\nonumber
\\ &&\int_{-1}^{1} du \int_{-1}^{1} dv \left |{u + v}\over 2 \right|^{2 y}
\int_{u}^{1} dt \int_{v}^{1} ds \left |{s + t}\over 2 \right|^{2 z},
\end{eqnarray}
where the path of integration over $z$ is chosen in such a
way that the poles
of the function $\Gamma(\omega -1 + z)$ lie to the left of
the path of integration and the poles of the function
$\Gamma(- z)$ lie to the
right of it (the same for the integration over $y$).

After the integration over $s$, $t$, $v$ and $u$ we have
 the following expression:

\begin{eqnarray}
\label{MB2}
C_{(13)(13)}&=& \alpha T^4 (4 L^{2})^{2 - 2 \omega}
{1 \over{[\Gamma(\omega-1)]^2}} {1 \over {(2 \pi i)^2}} \int_{-i \infty}^{+i
\infty} dy \Gamma(\omega -1 + y) \Gamma(-y) \left( {1 \over{- \beta^2 - i
\epsilon}}\right)^y \times\nonumber \\ &&\int_{-i \infty}^{+i \infty} dz
\Gamma(\omega
-1 + z) \Gamma(-z) \left( {1 \over{- \beta^2 - i \epsilon}}\right)^z \times
\nonumber
\\ &&{2^5 \over{(2 z + 1) (2 z + 2)}} \left[{1\over{(2 y + 1)
(2 y + 2)}} -
{1\over{(2 y + 1) (2 y + 2 z + 4)}}\right. \nonumber \\ && \left.
+{1\over{(2 y
+ 2 z + 3) (2 y + 2 z + 4)}} -  {{\Gamma(2 z + 3) \Gamma(2 y + 1)}
\over{\Gamma(2
z + 2 y + 5)}}\right].
\end{eqnarray}

Then one has  to integrate over $z$ and $y$; the integration contours have to
be suitably chosen: for instance, in the present example, in order to apply
Jordan's lemma, the integration paths  must be closed with
half-circles lying in the half planes ${\rm Re} z<0$ and   ${\rm Re} y<0$.
These integrations produce several double power series in the variable
$\beta^2$ with finite convergence radii, which are  particularly suited
for a large $T$ (small $\beta$) expansion.
For instance,  the last term in
the square bracket of  eq. (\ref{MB2}) gives the following contribution
\begin{eqnarray}
&&\alpha T^4 (4 L^{2})^{2 - 2 \omega}
{1 \over{[\Gamma(\omega-1)]^2}} {1 \over {(2 \pi i)^2}} \int_{-i \infty}^{+i
\infty} dy \Gamma(\omega -1 + y) \Gamma(-y) \left( {1 \over{- \beta^2 - i
\epsilon}}\right)^y
\nonumber \\ && \times \int_{-i \infty}^{+i \infty} dz
\Gamma(\omega
-1 + z) \Gamma(-z) \left( {1 \over{- \beta^2 - i \epsilon}}\right)^z
{2^5 \over{(2 z + 1) (2 z + 2)}} \left[
 -  {{\Gamma(2 z + 3) \Gamma(2 y + 1)}\over{\Gamma(2
z + 2 y + 5)}}\right]\nonumber \\
&&={{(2T)^{4-4\omega}}\over {\pi^{2\omega}}}(LT)^2
\Bigg\{e^{-2i\pi\omega}
{{\beta}^{-2}}(-\frac{1}{8}) \left({\pi \over{\sin(\pi\omega)}}\right)^2
\frac{\Gamma^2(\frac{3}{2}-\omega)}{\Gamma(\frac{1}{2})
\Gamma(\frac{9}{2} - 2 \omega) \Gamma(5 - 2\omega)} \nonumber \\
&& \times F_4(2 \omega - \frac{7}{2},2 \omega - 4, \omega -\frac{1}{2},
\omega -\frac{1}{2};\beta^2,\beta^2)
+ e^{-i\pi\omega} {\beta}^{1 - 2 \omega}
\frac{i}{4} {\pi \over{\sin(\pi\omega)}}
\frac{\Gamma(\frac{1}{2})\Gamma(\frac{3}{2}-\omega)\Gamma(\omega -
\frac{3}{2})}{\Gamma(3-\omega)\Gamma(\frac{7}{2}-\omega)}
\nonumber \\
&&\times F_4(\omega - \frac{5}{2},\omega - 2, \omega -\frac{1}{2},
\frac{5}{2}-\omega;\beta^2,\beta^2)
+ e^{-i\pi\omega} {\beta}^{2 - 2 \omega}
\frac{1}{4} {\pi \over{\sin(\pi\omega)}} \Gamma(\frac{1}{2})\nonumber\\
&& \times \sum_{l=0}^\infty \sum_{m=0}^\infty \frac
{\Gamma(m+1)\Gamma(l+m+\omega-\frac{3}{2})\Gamma(l+m+\omega-2)}
{\Gamma(l+\omega-\frac{1}{2})\Gamma(m+3-\omega)\Gamma(m+\frac{3}{2})}
\frac{(\beta^2)^l}{l!}
\frac{(\beta^2)^m}{m!}+{\beta}^{4 - 4 \omega}
\frac{\Gamma^2(\omega-\frac{3}{2})\Gamma^2(\frac{1}{2})}{4}\nonumber \\
&& + i {\beta}^{5 - 4 \omega} \Gamma(\frac{1}{2})\Gamma(\omega-2)
\Gamma(\omega-\frac{3}{2}) -\frac{{\beta}^{6 - 4 \omega}}{2}
\left[\frac{1}{2} \Gamma^2(\frac{1}{2})\Gamma(\omega-\frac{3}{2})
\Gamma(\omega-\frac{5}{2}) + \Gamma^2(\omega - 2)\right]\Bigg\}
\label{MB3}
\end{eqnarray}
with the notations as in \cite{erd}.

Repeating this procedure for each integral (A1)-(A7), 
one eventually recover eq.
(\ref{wmlg4}).

As a check of our calculations, we have explicitly verified the Abelian
exponentiation theorem. The sum of all the crossed diagrams, which are
proportional to $C_F^2- (1/2) C_FC_A$, behaves like  $(LT)^2 T^{4 - 4\omega}$,
and therefore is depressed in the large $T$ limit as long as $\omega >1$. This
means that only planar diagrams should contribute to the Abelian exponentiation
in the large $T$ limit. As a matter of fact this is in fact what happens:
there is a single planar
diagram that,
alone, provides the dominant term for the  Abelian exponentiation;
it is the one
in Fig. (2a). It can be  checked  that for such a diagram the leading
term in the large $T$ expansion is
$(-1/2\pi^{2\omega})((2L)^{4-4\omega}(C_FLT)^2
(\Gamma(\omega-3/2)\Gamma(1/2))^2$, which is precisely  the half of the square
of the corresponding leading order of the sum of the single exchange diagrams.

\section{bubble diagrams}
The complete one loop propagator (including ghost interaction) is given by
\begin{eqnarray}
D_{\mu\nu}^{(2)ab}& =& \delta^{ab} {g^2 C_A\over 16 \pi^{2\omega}}{(1 -
3\omega)(2-\omega)(3 - 2 \omega)\Gamma (1-\omega)\Gamma^2(\omega) \Gamma
(2\omega - 4)\over \Gamma(2 \omega)\Gamma(4 - \omega)(-x^2 + i
\epsilon)^{2\omega - 3}}\times\nonumber\\
&&\left[ {x_\mu x_\nu\over (- x^2 + i\epsilon)} - g_{\mu\nu}{2\omega - 5\over 2
(3 - 2\omega)}\right].
  \end{eqnarray}

Writing explicitly all the possible bubble diagrams $B_{ij}$, it is not
difficult to realize that  $B_{11}=B_{33}$, $B_{22}=B_{44}$, $B_{12}=B_{34}$,
 $B_{14}=B_{23}$. In addition, the two last pairs of bubbles are in turn equal
after a trivial change of variables, so that eq. (\ref{bubblelist}) follow.
The five independent bubbles are then
\begin{eqnarray}
B_{11}&=& \int_{-1}^1\!\! ds \int_{-1}^1 \!\! dt { (\sigma T^2)\over [-(s-t)^2
T^2 + i\epsilon]^{2\omega -3}}\left[{2\omega -1 \over 2 (3 -
2\omega)}\right]\ , \\
B_{22}&=& \int_{-1}^1\!\! ds \int_{-1}^1 \!\! dt { (\sigma L^2)\over [(s-t)^2
L^2 + i\epsilon]^{2\omega -3}}\left[{1 -2\omega \over 2 (3 -
2\omega)}\right]\ , \\
B_{13}&=& \int_{-1}^1\!\! ds \int_{-1}^1 \!\! dt { (-\sigma T^2)\over [-(s+t)^2
T^2 +4L^2+ i\epsilon]^{2\omega -3}}\left[{ (s+t)^2 T^2\over 4L^2 - (s+t)^2
T^2 + i \epsilon} - {2\omega - 5\over 2 (3 - 2\omega)} \right]\ ,\\
B_{24}&=& \int_{-1}^1\!\! ds \int_{-1}^1 \!\! dt { (-\sigma L^2)\over [(s+t)^2
L^2 -4T^2+ i\epsilon]^{2\omega -3}}\left[{ (s+t)^2 L^2\over (s+t)^2
L^2-4T^2 + i \epsilon} + {2\omega - 5\over 2 (3 - 2\omega)} \right]\ ,\\
B_{12}&=& \int_{-1}^1\!\! ds \int_{-1}^1 \!\! dt { \sigma L^2T^2
(t-1)(s+1)\over [(s+1)^2  L^2 - T^2 (t-1)^2+ i\epsilon]^{2\omega -2}}\ ,
\end{eqnarray}
where
\begin{equation}
\sigma={g^4 C_F C_A\over 16 \pi^{2\omega}}{(
3\omega-1)(2-\omega)(3 - 2 \omega)\Gamma (1-\omega)\Gamma^2(\omega) \Gamma
(2\omega - 4)\over \Gamma(2 \omega)\Gamma(4 - \omega)}
\end{equation}

Again, integration can be performed using Mellin-Barnes techniques leading to
\begin{eqnarray}
&&{\cal W}^{(2)}\!=\!{C_FC_A (2L)^{4 \! -\! 4\omega} (LT)^2 \Gamma^2 (\omega)
(3\omega
\! -\!1) \Gamma(1 \! -\!\omega)\Gamma (2\omega \! -\!2)\over 2 \pi^{2\omega}
\Gamma(2\omega)
\Gamma (4\! -\!\omega)} \times\nonumber\\
&&\left\{ e^{\! -\!2i\pi\omega}\beta^{4\omega \! -\!6}\left[ {(2\omega\! -\!
1)\over (4\omega \! -\!6)}
{1 \! -\! (4\omega \! -\!7) (1\! -\!\beta^2)^{4\! -\!2\omega} \! +\!
(4\omega \! -\!8)
_2F_1(2\omega\! -\!3,2\omega \! -\!7/2; 2\omega\! -\!5/2;\beta^2)\over
(2\omega \! -\!4)(4\omega
\! -\!7)}\right.\right.\nonumber\\
&&\left.\left. -{2\omega\! -\!1\over (4\omega \! -\!6)(2\omega\! -\!4)}
 [_2F_1(2\omega
\! -\!4,\! -\!1/2;1/2;\beta^2)\! -\!1] \! +\!{1\over 2\omega\! -\!3}
[_2F_1(2\omega
\! -\!3,\! -\!1/2;1/2;\beta^2)\! -\!1] \right.\right.\nonumber\\
&&\left.\left. +{1\over
(2\omega\! -\!4)(2\omega\! -\!3)} [(1\! -\!\beta^2)^{4\! -\!2\omega}
\! -\!1] \right]\right.\nonumber\\
&&\left. +e^{\! -\!2i\pi\omega} \beta^{4\omega \! -\!4} {2(2\omega
\! -\!3) _2F_1 (2\omega\! -\!2,
2\omega \! -\!5/2; 2\omega \! -\!3/2;\beta^2) \! -\! (4\omega\! -
\!5) (1\! -\!\beta^2)^{3\! -\!2\omega}\over
(2\omega \! -\!3)(4\omega \! -\!5)}\right. \nonumber\\
&&\left.+ i\beta{(\omega \! -\!3)\Gamma(1/2)\Gamma (2\omega \!
-\!7/2)\over \Gamma(2\omega
\! -\!2)} \! +\! \beta^2 {3\! -\!\omega\over(\omega\! -\!2)
(4\omega \! -\!7)}\right\}\ .
\label{bolleapp}
\end{eqnarray}
By performing the large $T$ limit in eq. (\ref{bolleapp}) one arrives at eq.
(\ref{bubble}).

\section{spider diagrams}
The spider diagrams are by far the most complicated to evaluate.
The sum of  the 6 inequivalent  spider diagrams, with the appropriate weights,
is given by

\begin{eqnarray}
\label{spider1app}
&&{\cal
W}^{(3)}=2(S_{124}+S_{123}+S_{122}+S_{144}+S_{112}+S_{114})\equiv\nonumber\\
&&{C_FC_A\Gamma(2\omega-2)L^2T^{6-4\omega}\over 32\pi^{2\omega}} \int_0^1
d\rho_1 \int_0^1d\rho_2\int_0^1d\rho_3\delta (1-\rho_1-\rho_2-\rho_3)
(\rho_1\rho_2\rho_3)^{\omega-2}\times\nonumber\\
&&\left\{ \int_{-1}^1\!\!ds_1\int_{-1}^1\!\!ds_2\int_{-1}^1\!\!ds_3
(\rho_1 (\rho_1-1) + \rho_2(\rho_2-1) - 2\rho_1\rho_2 + (\rho_2-\rho_1)s_3
\rho_3)\times \right.\nonumber\\
&&\left. [(\rho_1 \!-\!\rho_2 \!- \! s_3\rho_3)^2\!-\!\beta^2 (s_1
\rho_1 \!-\!s_2\rho_2 \!+\! \rho_3)^2\!-\!\rho_1(1\!-\!\beta^2 s_1^2)\!
-\!\rho_2 (1\!-\!\beta^2 s_2^2)\!
-\!\rho_3(s_3^2\!-\!\beta^2)\!+\!i\epsilon]^{2-2\omega}\right.\nonumber\\
&&\left.+\int_{-1}^1\!\!ds_1\int_{-1}^1\!\!ds_2\int_{-1}^1\!\!ds_3 (\rho_1
(\rho_1-1) + \rho_3(\rho_3-1) - 2\rho_1\rho_3 + (\rho_1-\rho_3)s_2
\rho_2)\times\right.\nonumber\\
&&\left. [(s_1\rho_1 \!-\!\rho_2 \!-\!
s_3\rho_3)^2\!-\!\beta^2 ( \rho_1 \!+\!s_2\rho_2 \!-\!
\rho_3)^2\!-\!\rho_1(s_1^2\!-\!\beta^2 ) \!-\!\rho_2 (1\!-\!\beta^2 s_2^2)
\!-\!\rho_3(s_3^2\!-\!\beta^2)\!+\!i\epsilon]^{2- 2\omega}\right.\nonumber\\
&&\left.+\int_{-1}^1 \!\! ds_1 \int_{-1}^{s_1}\!\! ds_2 \int_{-1}^{1} \!\!ds_3
(\rho_1 (\rho_1-1) - \rho_2(\rho_2-1) + (\rho_1-\rho_2)s_3
\rho_3)\times\right.\nonumber\\
&&\left. [(\rho_1 \!+\!\rho_2 \!+\! s_3\rho_3)^2\!-\!\beta^2
(s_1 \rho_1 \!+\!s_2\rho_2 \!-\! \rho_3)^2\!-\!\rho_1(1\!-\!s_1^2\beta^2 )
\!-\!\rho_2 (1\!-\!\beta^2 s_2^2)
\!-\!\rho_3(s_3^2\!-\!\beta^2)\!+\!i\epsilon]^{2-2\omega}\right.\nonumber\\
&&+\left.\int_{-1}^1\!\!ds_1\int_{-1}^{s_1}\!\!ds_2\int_{-1}^1\!\!ds_3 (\rho_1
(1-\rho_1) + \rho_2(\rho_2-1)  + (\rho_1-\rho_2)s_3 \rho_3)\right.
\times\nonumber\\
&&\left. [(\rho_1 \!+\!\rho_2 \!-\! s_3\rho_3)^2\!-\!\beta^2 (
s_1\rho_1 \!+\!s_2\rho_2 \!+\! \rho_3)^2\!-\!\rho_1(1\!-\!s_1^2\beta^2 )
\!-\!\rho_2 (1\!-\!\beta^2 s_2^2)
\!-\!\rho_3(s_3^2\!-\!\beta^2)\!+\!i\epsilon]^{2-2\omega}\right.\nonumber\\
&&+\left.\int_{-1}^1\!\!ds_1\int_{-1}^1\!\!ds_2\int_{-1}^{s_2}\!\!ds_3 (\rho_2
(1-\rho_2) + \rho_3(\rho_3-1)  + (\rho_2-\rho_3)s_1
\rho_1)\times\right.\nonumber\\
&& \left. [(\rho_1 \!+\! s_2\rho_2 \!+\!
s_3\rho_3)^2\!-\!\beta^2 ( s_1\rho_1 \!-\!\rho_2 \!-\!
\rho_3)^2\!-\!\rho_1(1\!-\! s_1^2\beta^2 ) \!-\!\rho_2 (s_2^2\!-\!\beta^2 )
\!-\!\rho_3(s_3^2\!-\!\beta^2)\!+\!i\epsilon]^{2- 2\omega}\right.\nonumber\\
&&+\left.\int_{-1}^1\!\!ds_1\int_{-1}^1\!\!ds_2\int_{-1}^{s_2}\!\!ds_3 (\rho_2
(\rho_2-1) + \rho_3(1-\rho_3) + (\rho_2-\rho_3)s_1
\rho_1)\right.\times\nonumber\\
&&\left. [(\rho_1 \!-\!s_2\rho_2 \!-\!
s_3\rho_3)^2\!-\!\beta^2 ( s_1\rho_1 \!+\!\rho_2 \!+\!
\rho_3)^2\!-\!\rho_1(1\!-\!s_1^2\beta^2 ) \!-\!\rho_2 (s_2^2\!-\!\beta^2)
\!-\!\rho_3(s_3^2\!-\!\beta^2)\!+\!i\epsilon]^{2-2\omega}\right\} \nonumber
\end{eqnarray}

The above integrals can be more conveniently grouped as
\begin{equation}
{\cal W}^{(3)}={C_FC_A(LT)^2(2 T)^{4-4\omega}\over \pi^{2\omega}}
e^{-2i\pi\omega} [I_1(\beta^2)+
I_2(\beta^2)+I_3(\beta^2)+I_4(\beta^2)]
\end{equation}
where
\begin{eqnarray}
I_1(\beta^2)& =& \Gamma(2\omega-3)
2^{4\omega-10}\int_0^1\! [d\rho]\int_{-1}^1\! [ds]
(\rho_1\rho_2\rho_3)^{\omega-2}
{\rho_1-\rho_2\over\rho_1+\rho_2}{\partial\over\partial s_3}\nonumber\\
&&[P_1^{3-2\omega} +\theta(s_1-s_2) P_2^{3-2\omega} + \theta(s_1-s_2)
P_3^{3-2\omega}]\nonumber\\
I_2(\beta^2)& =& \Gamma(2\omega-3)
2^{4\omega-10}\int_0^1\! [d\rho]\int_{-1}^1\! [ds]
(\rho_1\rho_2\rho_3)^{\omega-2} {\rho_2-\rho_1\over
\beta^2(\rho_1+\rho_2)}{\partial\over\partial s_3}\nonumber\\
 &&[P_4^{3-2\omega}
+\theta(s_1-s_2) P_5^{3-2\omega} + \theta(s_1-s_2) P_6^{3-2\omega}]\nonumber\\
I_3(\beta^2)& =& -\Gamma(2\omega-2) 2^{4\omega-7}\int_0^1\!
[d\rho]\int_{-1}^1\! [ds] (\rho_1\rho_2\rho_3)^{\omega-2}
{\rho_1\rho_2\over\rho_1+\rho_2}
P_1^{2-2\omega}\nonumber\\
I_4(\beta^2)& =& -\Gamma(2\omega-2) 2^{4\omega-7}\int_0^1\!
[d\rho]\int_{-1}^1\! [ds] (\rho_1\rho_2\rho_3)^{\omega-2}
{\rho_1\rho_2\over\rho_1+\rho_2}
P_4^{2-2\omega}
\end{eqnarray}

where $[ds]=ds_1ds_2ds_3$,
$[d\rho]=d\rho_1d\rho_2d\rho_3\delta(1-\rho_1-\rho_2-\rho_3)$ and
\begin{eqnarray}
P_1&\! =\!&(\rho_1 \! -\!\rho_2 \! +\! s_3\rho_3)^2\! -\!\beta^2 (s_1
\rho_1 \! -\!s_2\rho_2 \! -\! \rho_3)^2\! -\!\rho_1(1\! -\!s_1^2\beta^2 )
\! -\!\rho_2 (1\! -\!\beta^2 s_2^2)
\! -\!\rho_3(s_3^2\! -\!\beta^2)\! +\!i\epsilon\nonumber\\
P_2&\! =\!&(\rho_1 \! +\!\rho_2 \! +\! s_3\rho_3)^2\! -\!\beta^2 (s_1
\rho_1 \! +\!s_2\rho_2 \! -\! \rho_3)^2\! -\!\rho_1(1\! -\!s_1^2\beta^2 )
\! -\!\rho_2 (1\! -\!\beta^2 s_2^2)
\! -\!\rho_3(s_3^2\! -\!\beta^2)\! +\!i\epsilon\nonumber\\
P_3&\! =\!& (\rho_1 \! +\!\rho_2 \! -\! s_3\rho_3)^2\! -\!\beta^2 (s_1
\rho_1 \! +\!s_2\rho_2 \! +\! \rho_3)^2\! -\!\rho_1(1\! -\!s_1^2\beta^2 )
\! -\!\rho_2 (1\! -\!\beta^2 s_2^2)
\! -\!\rho_3(s_3^2\! -\!\beta^2)\! +\!i\epsilon\nonumber\\
P_4&\! =\!& (s_1\rho_1 \! -\!s_2\rho_2 \! -\!\rho_3)^2\! -\!\beta^2 (
\rho_1 \! -\!\rho_2 \! +\!s_3 \rho_3)^2\! -\!\rho_1( s_1^2\! -\!\beta^2 )
\! -\!\rho_2 (s_2^2\! -\!\beta^2 )
\! -\!\rho_3(1\! -\! s_3^2\beta^2)\! +\!i\epsilon\nonumber\\
P_5&\! =\!& (s_1\rho_1 \! +\!s_2\rho_2 \! +\!\rho_3)^2\! -\!\beta^2 (
\rho_1 \! +\!\rho_2 \! -\!s_3 \rho_3)^2\! -\!\rho_1( s_1^2\! -\!\beta^2 )
 \! -\!\rho_2 (s_2^2\! -\!\beta^2 )
\! -\!\rho_3(1\! -\! s_3^2\beta^2)\! +\!i\epsilon\nonumber\\
P_6&\! =\!& (s_1\rho_1 \! +\!s_2\rho_2 \! -\!\rho_3)^2\! -\!\beta^2 (
\rho_1 \! +\!\rho_2 \! +\!s_3 \rho_3)^2\! -\!\rho_1( s_1^2\! -\!\beta^2 )
\! -\!\rho_2 (s_2^2\! -\!\beta^2 )
\! -\!\rho_3(1\! -\! s_3^2\beta^2)\! +\!i\epsilon\nonumber
\end{eqnarray}

As already anticipated in the main text, the above integrals have not been
evaluated exactly. However, the leading power of $T$ is just the factor
$T^{6-4\omega}$ contained in the overall constant, as can be easily realized by
the fact that the integrals $I_1 ,\cdots ,I_4$ are finite when evaluated for
$\beta=0$. In turn, only $I_1 (\beta^2=0)$ and $I_3 (\beta^2=0) $ can be
evaluated analytically, whereas for $I_2(\beta^2=0)$ and $I_4(\beta^2=0)$
we have only an expansion around $\omega=1$ that, however, is just what
we need. The results are
\begin{eqnarray}
I_1(\beta^2=0)&=& {\Gamma(2\omega-2)\over
3-2\omega}\left[{\Gamma(\omega)\Gamma(\omega+1)\over \omega
(\omega-2)^2\Gamma(2\omega-1)}-{\pi\over
(2\omega-4)\sin(\pi\omega)}\right]\nonumber\\
I_2(\beta^2=0)&=&-{1\over4(\omega-1)^2}+ {\gamma-1/2\over
2(\omega-1)}\nonumber\\
&&+{17\over 4}+{1\over2}\gamma(1-\gamma) + {7\over 24}\pi^2 - \pi^2 \log2 -
{3\over 2}\zeta(3) +{\cal O}(\omega -1)\nonumber\\
I_3(\beta^2=0)&=&
{2\over\Gamma(5-2\omega)}\left[\Gamma(2\omega-2)\Gamma(1-\omega)
\Gamma(3-\omega)+{\pi\over\sin(\pi\omega)}\times\right.\nonumber\\
&&\left.\sum_{n=1}^\infty {1\over n!}\left({\Gamma
(2\omega-2+n)\Gamma(3+n-\omega) \over
(2n+1)\Gamma(n+\omega)}-{\Gamma(n+\omega-1)\Gamma(4-2\omega +n)
\over(2n+3-2\omega)\Gamma(2-\omega+n)}\right)\right]\nonumber\\
I_4(\beta^2=0)&=&-{1\over 2(\omega-1)^2} +{\gamma -4\over (\omega
-1)}+\nonumber\\
&&+8\gamma -\gamma^2 - 22 +{\pi^2\over 12} +\pi^2\log2 +{3\over 2}\zeta(3)
+{\cal O} (\omega -1)
 \end{eqnarray}
By expanding $I_1$ and $I_3$ one can easily get eq. (\ref{ragnilaurent}).

\vfill\eject

\begin{figure}
\caption{Parametrization of the closed  rectangular loop $\gamma$ in four
segments $\gamma_i$.}
\label{fig1}
\end{figure}
\begin{figure}
\caption{Example of non-crossed and crossed diagrams.}
\label{fig2}
\end{figure}
\begin{figure}
\caption{Examples of crossed diagrams; they are labelled as $C_{(13)(13)}$,
$C_{(11)(11)}$ and $C_{(13)(11)}$, respectively}
\label{fig3}
\end{figure}
\begin{figure}
\caption{The three crossed diagrams that are unrelated to other diagrams through
symmetry relations; they are  $C_{(13)(13)}$,
$C_{(24)(24)}$ and $C_{(13)(24)}$. }
\label{fig4}
\end{figure}
\begin{figure}
\caption{Examples of bubble diagrams. They are labelled as $B_{11}$, $B_{13}$
and $B_{34}$,
respectively.}
\label{fig5}
\end{figure}
\begin{figure}
\caption{Examples of spider diagrams. They are labelled as $S_{123}$, $S_{112}$,
respectively.}
 \label{fig6}
\end{figure}

\vfill
\eject

\begin{picture}(400,620)(33,0)
\leavevmode
\epsfbox{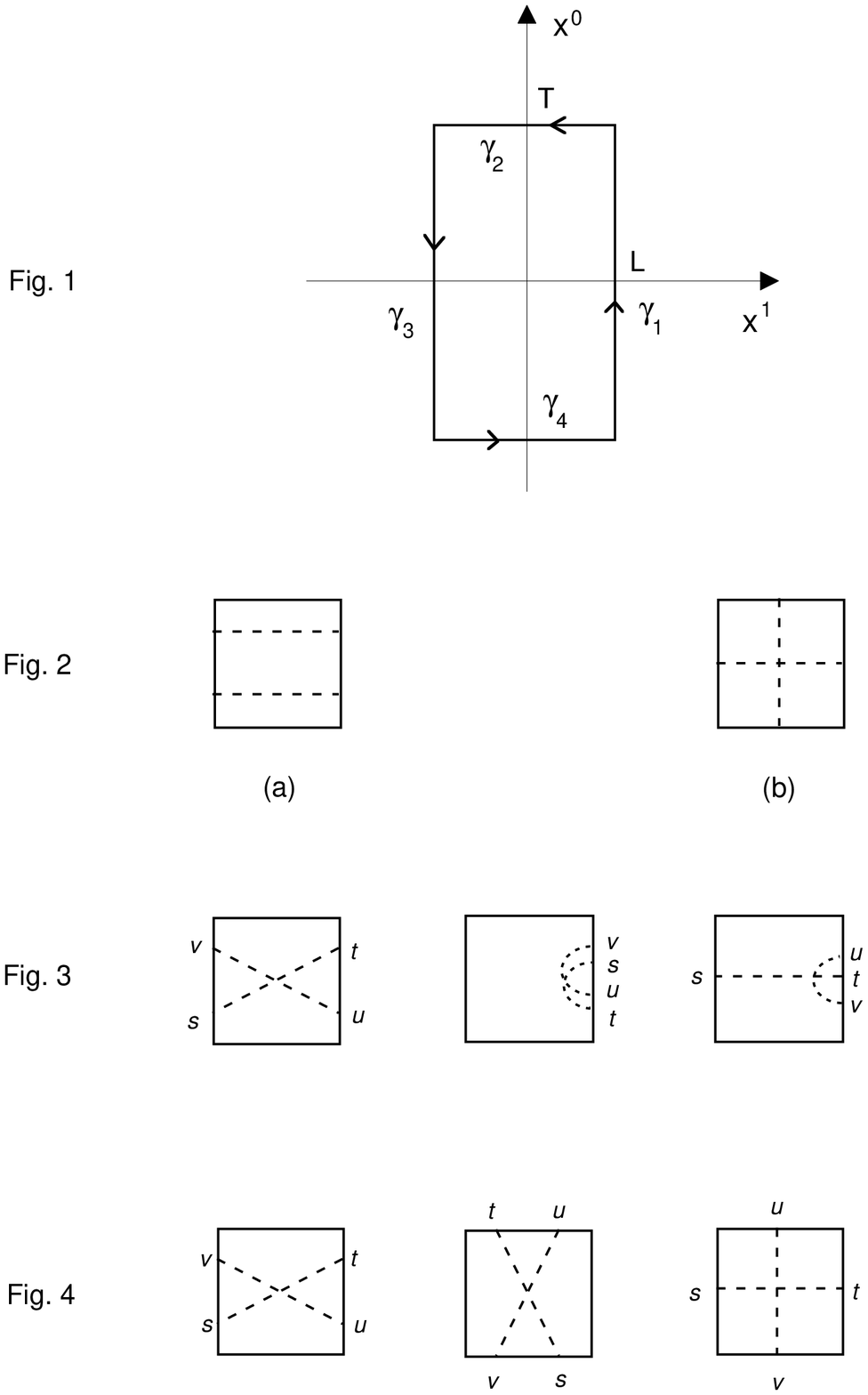}
\end{picture}

\vfill\eject

\begin{picture}(400,180)(33,0)
\leavevmode
\epsfbox{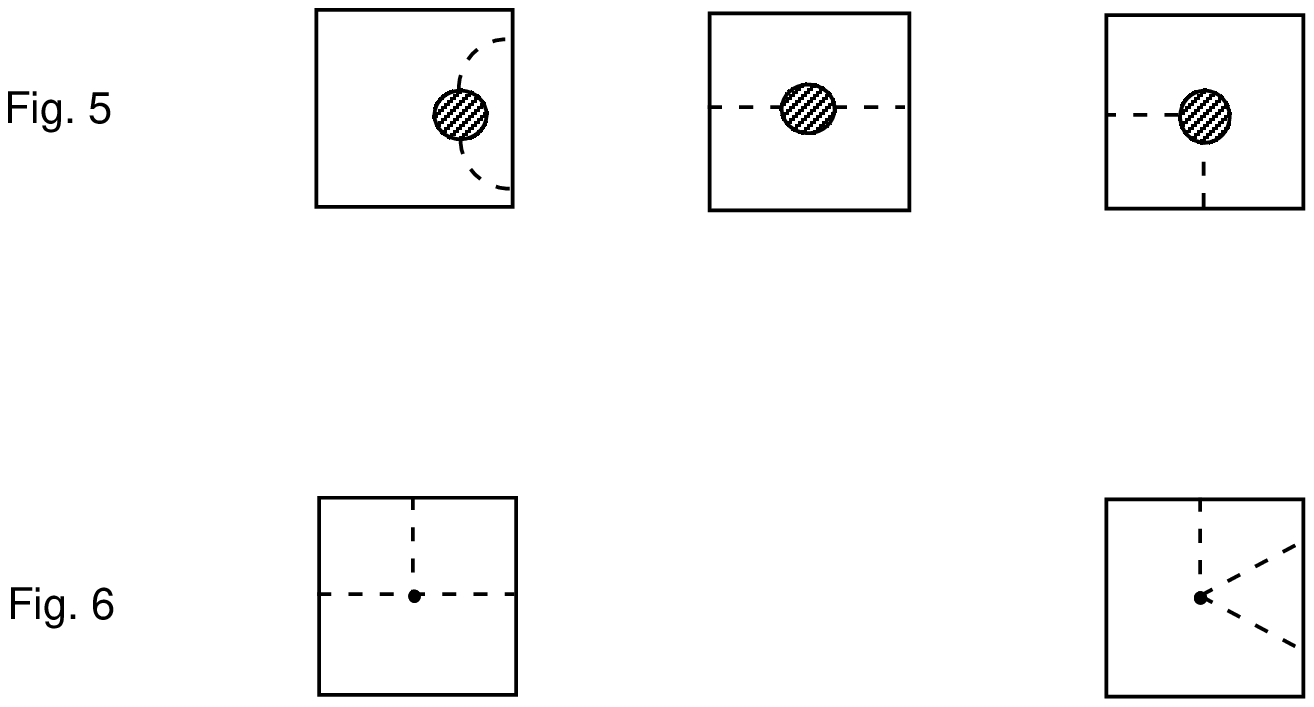}
\end{picture}


\begin{references}

\bibitem{poly72}{A. M. Polyakov, Phys. Lett. \underbar{82B}, 247 (1972);\\
J. B. Kogut and L. Susskind, Phys. Rev. \underbar{D11}, 395 (1975).}

\bibitem{fish}{W. Fishler, Nucl. Phys. \underbar{B129}, 157 (1977).}

\bibitem{wils74}{K. Wilson, Phys. Rev. \underbar{D10}, 2445 (1974);\\
L.S. Brown and W.I. Weisberger, {\it ibid} \underbar{20}, 3239 (1979).}

\bibitem{hoof74}{ G. 't Hooft, Nucl. Phys. \underbar{B75}, 461
(1974).}

\bibitem{call76}{ C.G. Callan, N. Coote and D.J. Gross, Phys.Rev.
\underbar{D 13}, 1649 (1976).}

\bibitem{Wu}{ T.T. Wu, Phys. Lett. \underbar{71B}, 142
(1977).}

\bibitem{ML}{S. Mandelstam, Nucl. Phys. \underbar{B213}, 149 (1983);\\
G. Leibbrandt, Phys. Rev. \underbar{D29}, 1699 (1984).}

\bibitem{Bas5}{A. Bassetto, M. Dalbosco, I. Lazzizzera and R. Soldati,
Phys. Rev. \underbar{D31}, 2012 (1985).}

\bibitem{Bas3}{A. Bassetto, M. Dalbosco and R. Soldati, Phys. Rev.
\underbar{D36}, 3138 (1987).}

\bibitem{Bas4}{ A. Bassetto, G. Nardelli and R. Soldati, {\it Yang--Mills
theories in algebraic non covariant gauges} (World Scientific,
Singapore, 1991).}

\bibitem{Bas1}{A. Bassetto, F. De Biasio and L. Griguolo,
Phys. Rev. Lett. \underbar{72}, 3141 (1994).}

\bibitem{Korc}{ I.A. Korchemskaya and G.P. Korchemsky,
Phys. Lett. \underbar{B 287}, 169 (1992).}

\bibitem{Bas2}{A. Bassetto, I.A. Korchemskaya, G.P. Korchemsky and G.
Nardelli, Nucl. Phys. \underbar{B 408}, 62 (1993).}

\bibitem{ALTE}{S. Caracciolo, G. Curci and P. Menotti, Phys. Lett.
\underbar{113B}, 311 (1982);\\
J.P. Leroy, J. Micheli and G.C. Rossi, Nucl. Phys. \underbar{B232},
511 (1984).}
 
\bibitem{Bas7}{A. Bassetto, D. Colferai and G. Nardelli, Nucl. Phys.
\underbar{B501}, 227 (1997) and E. {\it ibid} \underbar{507}, 746 (1997).}

\bibitem{Stau}{M. Staudacher and W. Krauth, Phys. Rev. \underbar{D57},
2456 (1998).}

\bibitem{Fre}{J. Frenkel and J. C.Taylor, Nucl. Phys. \underbar{B246},
 231 (1984).}

\bibitem{doug}{E. Witten, Commun. Math. Phys. \underbar{141}, 153 (1991);\\
M.R. Douglas, K. Li and M. Staudacher, Nucl.Phys.
\underbar{B420}, 118 (1994).}

\bibitem{boul}{M.R. Douglas and V.A. Kazakov, Phys. Lett. \underbar{B319},
219 (1993);\\
D.V. Boulatov, Mod. Phys. Lett. \underbar{A9}, 365 (1994);\\
D.J. Gross and A. Matytsin, Nucl. Phys. \underbar{B429}, 50 (1994).\\
We thank L. Griguolo for calling our attention
to these papers.}

\bibitem{Basgri}{A.Bassetto and L. Griguolo, hep-th/9806037.}
\bibitem{erd}{{\it Higher Trascendental Functions}, Vol. 1, Ed. A. Erd\'elyi,
McGraw-Hill , New York 1953.}
\end{references}
\end{document}